\documentclass[ 
superscriptaddress,
showpacs,
showkeys,
preprintnumbers,
 amsmath,amssymb,nofootinbib,
 aps,
prd, twocolumn,  
floatfix,
usenames,
dvipsnames
]{revtex4-1}

\usepackage{epsfig}
\usepackage{graphicx}
\usepackage{dcolumn}
\usepackage{bm}
\usepackage{hyperref}
\usepackage{natbib}
\usepackage{comment}
\usepackage{xcolor}
\usepackage{enumerate}

\DeclareMathOperator*{\maxi}{max}
\newcommand{\yr}{\, \text{yr}}
\newcommand{\SNR}{\text{SNR}}
\newcommand{\M}{\text{M}}
\newcommand{\PN}{{\text{PN}}}
\newcommand\msun{{{\rm M}_\odot}}
\newcommand{\uvec}[1]{\hat{\bm{#1}}}
\newcommand{\dvec}[1]{\dot{\bm{#1}}}
\newcommand{\duvec}[1]{\dot{\hat{\bm{#1}}}}

\begin{document}

\preprint{APS/123-QED}

\title{Post-Newtonian phase accuracy requirements for stellar black hole binaries with LISA}

\author{Alberto Mangiagli}
 \email[E-mail: ]{a.mangiagli@campus.unimib.it}
\affiliation{%
Department of Physics G. Occhialini, University of Milano - Bicocca, Piazza della Scienza 3, 20126 Milano, Italy
}%
\affiliation{%
National Institute of Nuclear Physics INFN, Milano - Bicocca, Piazza della Scienza 3, 20126 Milano, Italy
}%
\author{Antoine Klein}
\affiliation{Institut d'Astrophysique de Paris, CNRS \& Sorbonne
 Universit\'es, UMR 7095, 98 bis bd Arago, 75014 Paris, France}
\affiliation{School of Physics and Astronomy, University of Birmingham, Edgbaston,
Birmingham B15 2TT, United Kingdom}

\author{Alberto Sesana}
\affiliation{School of Physics and Astronomy, University of Birmingham, Edgbaston,
Birmingham B15 2TT, United Kingdom}

\author{Enrico Barausse}
\affiliation{Institut d'Astrophysique de Paris, CNRS \& Sorbonne
 Universit\'es, UMR 7095, 98 bis bd Arago, 75014 Paris, France}

\author{Monica Colpi}
\affiliation{%
Department of Physics G. Occhialini, University of Milano - Bicocca, Piazza della Scienza 3, 20126 Milano, Italy
}%
\affiliation{%
National Institute of Nuclear Physics INFN, Milano - Bicocca, Piazza della Scienza 3, 20126 Milano, Italy
}%

\date{\today}%

\begin{abstract}
The Laser Interferometer Space Antenna (LISA) will observe black hole binaries of stellar origin during their gravitational wave inspiral, months to years before coalescence. Due to the long duration of the signal in the LISA band, a faithful waveform is necessary in order to keep track of the binary phase. This is crucial to extract the signal from the data and to perform an unbiased estimation of the source parameters. We consider Post-Newtonian (PN) waveforms, and analyze the
 PN order needed to keep the bias caused by the PN approximation
 negligible relative to the statistical parameter estimation error, as a function of the source parameters. 
By considering realistic  population models, we  conclude that for $\sim 90\%$ of the stellar black hole binaries detectable by LISA, waveforms at low Post-Newtonian (PN) order (PN $\le 2$) are sufficiently accurate for an unbiased recovery of the source parameters. Our results provide a first estimate of the trade-off between waveform accuracy and information recovery for this class of LISA sources.
\end{abstract}

\pacs{
 04.30.-w, 
 04.30.Tv 
}

\keywords{LISA - Post-Newtonian theory}

\maketitle

\section{\label{sec:intro} Introduction}
The historical detection of the gravitational wave (GW) from the source GW150914 marked the dawn of a new era of  scientific exploration \cite{PhysRevLett.116.061102}. 
The GW signal from GW150914 is compatible with the coalescence of two stellar-origin black holes with component masses $36^{+5}_{-4} \, \M_{\odot}$ and $29^{+4}_{-4} \, \M_{\odot}$ at redshift $z = 0.09^{+0.03}_{-0.04}$. 
 In the following months, the Advanced LIGO interferometers \cite{0264-9381-27-8-084006} detected more stellar-origin black-hole binaries (SBHBs), namely GW151226 \cite{PhysRevLett.116.241103}, GW170104 \cite{PhysRevLett.118.221101} and GW170608 \cite{2041-8205-851-2-L35}. On August 14, 2017 the LIGO and Advanced Virgo interferometers, together, detected the
signal from the coalescence of two black holes (GW170814) \cite{PhysRevLett.119.141101}. The joint observation enabled an accurate sky localization of the source, with a $90\%$ credible region of $60 \, \text{deg}^2$, which allowed testing the tensor nature of GWs. Three days later, the GW signal from the coalescence of two neutron stars was observed, together with its electromagnetic counterpart \cite{PhysRevLett.119.161101, GBM:2017lvd}. 
Not only did these detections provide exquisite tests of General Relativity in the dynamical strong-field limit \cite{PhysRevLett.116.221101} as well as tests of the propagation of GWs over cosmic distances, but they also proved the existence of a rich variety of astrophysical black holes and neutron stars forming in binary systems \cite{2041-8205-818-2-L22}. 
The high incidence (in the five events) of ``heavy'' stellar black holes with masses in excess of $20\, \M_\odot$, and the high merger rates inferred by LIGO-Virgo imply 
the existence of a large population of fairly massive SBHBs accessible also to space-based interferometers, as pointed out in \cite{PhysRevLett.116.231102}.

LIGO and Virgo are sensitive to GW signals in the frequency range between $[10, \, 10^3] \, \text{Hz}$, as at frequencies lower than 10 Hz they are limited by the Newtonian ground noise. As a consequence, they are only able to observe the very late inspiral, merger and ring-down phases of SBHB coalescences, lasting typically less than one second \cite{PhysRevLett.116.241102}.  By contrast, the space mission LISA (Laser Interferometer Space Antenna) will detect GW signals in the frequency domain from below $0.1 \,\rm {mHz}$ to above $0.1 \, \text{Hz}$ \cite{2017arXiv170200786A}. SBHBs emit this low frequency radiation during their early inspiral, detectable  from weeks  to years before they merge in LIGO/Virgo band. Furthermore, many more systems will be observed as slowly inspiralling sources around 5-10 mHz 
hundreds of years before their final coalescence \cite{PhysRevLett.116.231102,Kyutoku:2016ppx}. This new population might prove useful in a number of astrophysical and cosmological contexts. For example, eccentricity measurements which are possible in the LISA band \cite{PhysRevD.94.064020} can be used to discriminate among different astrophysical formation pathways \cite{Nishizawa:2016eza,Breivik:2016ddj,2018arXiv180406519S,2018arXiv180506194D}. With a LISA mission duration time of 4 years extendable to 10 years, a population of SBHBs should be detected with GW signals lasting up to millions of inspiral cycles in band.
Therefore, an accurate description of the waveform is necessary to track the signal during this early inspiral phase. A mismodeled waveform translates in a phase difference with the true signal. Such a systematic error
produces biased results when estimating the source parameters~\cite{PhysRevD.91.124062,PhysRevD.93.064001, Abbott:2016wiq, PhysRevD.96.124041, PhysRevD.89.103012, PhysRevLett.112.101101, PhysRevD.63.082005, PhysRevD.88.062001}, and even cause residual power to be left in the data when the signal is subtracted (which could hinder further detections). For larger phase differences, even the detection of the signal may be in potential jeopardy.

Since both parameter estimation and detection require cross-correlating rapidly a stretch of data with a bank of template signals covering the full possible parameter space of the source, a delicate balance must be found between the waveform's accuracy and its production speed (which depends in turn on the waveform's complexity). For this reason, 
in this work  we address the question of how accurate a waveform has to be in order to follow a SBHB signal for a period of months or years in the LISA band. During the inspiral phase, the SBHB signal can be modeled using the Post-Newtonian (PN) formalism \cite{Blanchet2014}, and thus we need to assess the lowest PN order necessary to produce a faithful waveform (i.e. a waveform suitable for parameter estimation; note that this imposes stricter requirements than simple detection).

In order to estimate the minimum waveform accuracy requirements, we perform different sets of simulations, exploring plausible values for the mass ratio, total binary mass  and coalescence time  under the assumption that the binaries follow circular orbits and using suitable models (calibrated with the observed LIGO/Virgo merger rate) for the underlying astrophysical population. 

This paper is organized as follows. In Section \ref{sec:signal_part} we review the
basics of GW data analysis, and in particular the concepts of matched filtering, optimal signal-to-noise ratio (SNR), overlap and waveform faithfulness. 
In Section \ref{sec:waveform}, we review the PN black-hole  waveforms that we will use for this study.  In Section \ref{sec:results}, we present the results of our analysis, computing the faithfulness between mock true signals and templates with different PN contributions. We apply our framework to mock SBHB catalogs extracted from population models consistent with current LIGO/Virgo constrains in Section \ref{sec:results_pop_model}, and summarize our findings in Section \ref{sec:discussion_conclusion}.

\section{\label{sec:signal_part} Matched filtering, optimal signal-to-noise ratio, and faithfulness}

Estimating the parameters of a GW source involves cross-correlating the detector's output (which is comprised of the GW signal plus the instrumental noise of the interferometer) with a bank of templates. The latter depend on the intrinsic parameters of the source (masses, spins, initial separation, and eccentricity if present) as well as on the extrinsic  ones (orientation relative to the detector, polarization angle, sky position, distance, time and phase at coalescence), and are typically obtained by solving the Einstein equations of General Relativity.\footnote{Phenomenological deviations from General Relativity are however introduced when GW data are used to test the correctness of the theory~\cite{PhysRevLett.116.221101}.}
Because of their high computational cost, exact numerical solutions to these equations, however, are not sufficient by themselves to cover the whole parameter space~\cite{PhysRevX.6.041014}, due to its large dimensionality. Therefore, several families of approximate waveforms have been put forward, depending on the particular class of sources under scrutiny. This may in principle result in systematic deviations (``mismodeling biases'') of the inferred parameters from their true values, on top of the statistical errors due to the detector's noise. In the following we will briefly review the effect of these systematic errors on the waveforms, and focus in particular on their impact on detection and parameter estimation. In doing so we will introduce the concepts of optimal SNR, overlap, and faithfulness. We refer to the original literature on these topics (e.g.~\cite{PhysRevD.49.2658,PhysRevD.53.6749,PhysRevD.52.605,PhysRevD.57.885,PhysRevD.78.124020}) as well as to textbook presentations such as \cite{Maggiore:1900zz} for more details.

Consider a detector output of the form $s(t) = h(t) + n(t)$, where $h(t)$ represents the GW signal and $n(t)$ the detector noise. We assume the noise to be stationary and Gaussian with zero mean. We define the noise spectral density $S_n(f)$ of the interferometer as
\begin{equation}
 \langle \tilde{n}^{*}(f)\tilde{n}(f') \rangle = \frac{1}{2} \delta(f -f')S_n(f),
\end{equation}
where $\tilde{n}(f)$ denotes the Fourier transform of $n(t)$ 
and $\tilde{n}^{*}(f)$ its complex conjugate. The bracket $\langle \dots \rangle$ represents an ensemble average (which can be approximated with stationary noise by a time average, as a consequence of the ergodic theorem). 

The SNR can be computed as the ratio $S/N$ between the signal and noise amplitudes $S$ and $N$. 
To compute these amplitudes, let us first define the projection $\hat{s}$ of the time series $s(t)$ on a filter function $K(t)$, i.e. $\hat{s}=\int_{-\infty}^{+\infty} s(t) K(t) dt$. We can then define~\cite{Maggiore:1900zz} $S$ as the average of $\hat{s}$, i.e. $S=\langle\hat{s}\rangle=\int_{-\infty}^{+\infty} h(t) K(t) dt$, and $N$ as its rms value when no signal is present, i.e. $N^2=\langle\hat{s}^2\rangle\vert_{h=0}$.

Using the stationarity and Gaussianity of the noise, and defining the internal product
\begin{equation}
\label{eq:scalar_product_signal_space}
(a|b)  = 4 \,  \text{Re}  \int_{0}^{+\infty}df \, \frac{\tilde{a}^*(f)\tilde{b}(f)}{S_n(f)}
\end{equation}
between two real functions $a(t)$ and $b(t)$, one can easily show that~\cite{Maggiore:1900zz}
\begin{equation}
\text{SNR} =\frac{S}{N}= \frac{(u\,|\,h)}{(u\,|\,u)^{1/2}}\,,
\end{equation}
with $\tilde{u}(f)=\tilde{K}(f)S_n(f)/2$. From this expression, it is clear that the optimal function $K$ that can be used to
filter the detector's output is given by $K(f)\propto \tilde{h}(f)/S_n(f)$, which yields the optimal (maximum) SNR
\begin{equation}
\text{SNR}_{\rm opt} ={(h\,|\,h)^{1/2}}\,.
\end{equation}
However, in practice, the filter function will be constructed from a template bank, no element of which will in general reproduce the signal $h(t)$ exactly because of the systematic mismodeling mentioned above. Therefore, if one takes $K(f)\propto \tilde{h}_{t}(f)/S_n(f)$, where $h_t$ is a given template, then the SNR will be
\begin{equation}
\text{SNR} = \frac{(h\,|\,h_{{t}})}{(h_t\,|\,h_t)^{1/2}}=\mathcal{O}(h,h_{t}) \text{SNR}_{\rm opt}\,,
\end{equation}
where we have introduced the  overlap function
\begin{equation}
\mathcal{O}(h,h_{t}) \equiv \frac{(h|h_t)}{\sqrt{(h|h)(h_t|h_t)}}\,.
\end{equation}
This function therefore quantifies the reduction in SNR due to the use of a sub-optimal template.

Clearly, if the SNR reduction due to the template mismodeling is too large, it could hinder detection of the signal. A measure of the SNR reduction relevant for detection is given by the effectualness, which is defined as the maximum overlap function that can be obtained between a signal and the {\it whole} template bank (i.e.~it is the overlap maximized over all possible template parameters).
In practice, this means that we require that at least {\it one} template matches sufficiently well the signal, irrespective of the template parameters. However, this condition is still too loose for parameter estimation, because for the latter we would rather need the template best matching the signal to correspond to source parameters ``not too far'' from the true ones. More precisely, a useful measure of the performance of a template family for parameter estimation is  given by the  \emph{faithfulness}, i.e.~the overlap maximized only over the coalescence time $t_c$ and phase $\phi_c$ of the template, while adopting the true values for the other template parameters:
\begin{equation}
\label{eq:faithfulness}
F \equiv \maxi_{t_c,\phi_c}\frac{(h_t|h)}{\sqrt{(h_t|h_t)(h|h)}}.
\end{equation}

The threshold value of the faithfulness depends on the SNR.
Since a mismodeled waveform introduces systematic errors in the parameter estimation, a reasonable requirement is to lower these errors below the statistical ones resulting from instrumental noise. The latter can be estimated from the covariance matrix, which is simply the inverse of 
the Fisher matrix $\Gamma_{ij}=(\partial_ih_t|\partial_j h_t)$ (where $i,j$ denote the waveform's parameters) in the high-SNR regime.
In particular, in this regime the statistical errors decrease as $\SNR^{-1}$ \cite{PhysRevD.77.042001}, while the systematic errors are SNR-independent. 

In the high SNR regime, the expectation value of the faithfulness under the effect of statistical errors alone (i.e. resulting from the detector's noise and not from waveform mismodeling) in ~\cite{PhysRevD.95.104004} is given by 
\begin{equation}
\label{eq:expected_F}
\langle F\rangle \approx 1- \frac{D-1}{2 \, \text{(SNR)}^2}\,,
\end{equation}
where $D$ is the number of parameters describing the template. In order to allow for an unbiased parameter estimation, the faithfulness of the templates needs  to be less than the threshold given by Eq.~\eqref{eq:expected_F}. Similar requirements on $F$ were obtained also in \cite{PhysRevD.78.124020} as conditions on the amplitude and phase of the waveforms. 

\section{\label{sec:waveform} Waveform Generation}
The coalescence process of a SBHB consists of three different phases: \emph{inspiral}, \emph{merger}, \emph{ringdown}. In this study, we consider the inspiral phase only, as LISA can only detect the very early inspiral of a SBHB.
During this phase, when the binary contracts adiabatically and orbital velocities $v$ are much smaller than the speed of light $ c$, the equations of motion (EOM) are solved within the PN formalism, which consists of a series expansion in powers of $v/c$ or, equivalently, a series expansion in powers of the dimensionless gravitational potential.
At present, the EOM are known and can be solved up to $3.5$PN order if the binary components are not spinning or if their spins are aligned/anti-aligned with the binary angular momentum, i.e. neglecting precession effects. If precession effects are taken into account, the EOM are known up to $2.5$PN order, even if a closed form solution is still missing.

For this study, we use a Fourier-domain precessing waveform derived in~\cite{PhysRevD.90.124029}, augmented to 8PN non-spinning order at leading order in the mass ratio $q\equiv m_2/m_1\leq1$~\cite{Fujita:2012,Fujita:2015}.
We use a shifted uniform asymptotic (SUA) description of the gravitational wave signal produced by a generically spinning binary system in LISA. This method, being based on a
PN description of the orbital evolution, is particularly suited to describe SBHB systems as observed by LISA.
We use geometric units where $G=c=1,$ so that the dimensionless parameter $v = \omega^{1/3}$ is our PN expansion parameter, where $\omega$ is the orbital frequency.  We  numerically evolve the following equations of motion:
\begin{align}
 M \dot{\phi} &= v^3, \label{eq:phidot} \\
 M \dot{v} &= v^9 \sum_{n = 0}^{16} a_n v^n, \label{eq:vdot} \\
 M \duvec{L} &= -v^6 \left( \bm{\Omega}_1 + \bm{\Omega}_2 \right), \label{eq:Ldot}\\
 M \dvec{s}_1 &= \frac{m_2}{M} v^5 \bm{\Omega}_1, \\
 M \dvec{s}_2 &= \frac{m_1}{M} v^5 \bm{\Omega}_2, \label{eq:s2dot}
\end{align}
where $\phi$ is the orbital phase of the system, $\uvec{L}$ is the unit vector parallel to the orbital angular momentum $\bm L$, $\bm{s}_i = \bm{S}_i/m_i$ are the individual reduced spins. 
Equation \eqref{eq:phidot} contains as key timescale the orbital time ${\cal O}(v^{-3})$,  equation \eqref{eq:vdot} the radiation reaction timescale ${\cal O}(v^{-8})$, while equations \eqref{eq:Ldot}-\eqref{eq:s2dot} describe 
the relativistic precession occurring on a timescale ${\cal O}(v^{-5}$).
The vectors $\bm{\Omega}_i$ are defined in \cite{PhysRevD.90.124029} and  are orthogonal to $\bm{s}_i$, while $\bm{\Omega}_1+\bm{\Omega}_2$ is orthogonal to $\bm{L}$.
The coefficients  $a_n$ (with $0 \leq n \leq 7$) can also be found in~\cite{PhysRevD.90.124029} and are expressed in terms of powers of the symmetric mass ratio $\nu=q/(1+q)^2$.
They also contain terms related to both spin-orbit and spin-spin precession.

  In the following, we will refer to a waveform constructed from these equations (Eq.~\eqref{eq:vdot}) truncated at order $v^{9+2N}$ as an ``$N$-PN waveform''. In this, we follow the traditional PN order counting where the first non-vanishing term in an expression is called ``Newtonian'' or ``leading PN order'', and each further factor of $v^2$ adds one PN order to the term it multiplies. 
  Our expressions for the vectors $\bm{\Omega}_i$ are valid at leading PN order for the terms quadratic in the spins, and at 2PN
order (i.e. ${\cal O}(v^{4}$)
beyond the leading PN order) for the terms linear in the spins.

The constants $a_n$ are valid for generic mass ratios and spin configurations at 2PN order, and are valid at linear order in the spins at 3.5PN order. Note that the 4PN non-spinning term was recently computed~\cite{PhysRevD.95.044026,PhysRevD.97.044023,PhysRevD.97.044037}, and will be included in further studies.
While the constants $a_n$ are currently not know for generic mass ratios up to 8PN order, in order to simulate a high PN order frequency evolution equation and compare waveforms at high PN orders, we use a hybrid approach to construct $a_n$ for $n \geq 8$.

We first construct an 8PN relation between the orbital energy and the orbital frequency, $E(v),$ by including all generic terms up to 3.5PN~\cite{Blanchet2014}, and adding subsequent PN terms in the extreme mass ratio limit by expanding the relation for test particle circular orbits in a Schwarzschild spacetime~\cite{Fujita:2012}:
\begin{align}
E &= \frac{1 - 2 v^2}{\sqrt{1 - 3v^2}}. \label{eq:Eofv}
\end{align}
While the binding energy is known at the linear order in the mass ratio at 6PN order~\cite{LeTiec-2011,LeTiec-2011-2}, using this more accurate result would not add much complexity to the EOM, and therefore we choose the simpler approximate result \eqref{eq:Eofv}.

We then perform the same construction to get the gravitational wave luminosity $\mathcal{L}(v)$, by including all generic terms at 3.5PN by leading order~\cite{Blanchet2014}, and subsequent PN terms in the extreme mass ratio limit from~\cite{Fujita:2012}. We then use these two hybrid quantities to construct a simulated frequency evolution equation at 8PN beyond leading order by expanding the balance equation:
\begin{align}
\frac{dv}{dt} &= \frac{\mathcal{L}(v)}{dE/dv}.
\end{align}

To build the different waveforms that we compare in this work, we solve the equations of motion~(\ref{eq:phidot}-\ref{eq:s2dot}), truncating Eq.~\eqref{eq:vdot} at different PN orders, using a fifth order Cash-Karp Runge-Kutta method~\cite{Cash:1990}. Using those solutions, we then construct the LISA response to a passing GW in the time-domain as
\begin{align}
h(t) &= \sum_{n=1}^7 H_n(t) e^{-i n(\phi + \phi_T)},
\end{align}
where $H_n(t)$ are wave amplitudes for different GW harmonics that can be found at 2.5PN order for non-spinning systems in~\cite{Arun:2004,Arun:2004:erratum} and contain the detector response to the GW polarizations in the low-frequency approximation~\cite{Cutler:1997}, and $\phi_T$ is the Thomas precession phase accounting for the precession of the orbital plane in the detector frame and satisfies
\begin{align}
\dot{\phi}_T &= \frac{\cos \iota}{1 - \cos^2 \iota} \left(\uvec{L} \times \uvec{N} \right) \cdot \duvec{L},
\end{align}
where $\iota = \arccos(\uvec{L} \cdot \uvec{N})$ is the inclination angle, and $\uvec{N}$ is the line-of-sight vector from the detector to the source.

We then use those results in a SUA transform to compute the gravitational wave signal, as in~\cite{PhysRevD.90.124029}:
\begin{align}
 \tilde{h}(f) &= \sum_{n = 1}^7 \sqrt{2\pi} T_n e^{i(2\pi f t_n - n \phi(t_n) - \pi/4)} \nonumber\\
 &\times \sum_{k = -3}^3 b_{k} \mathcal{H}_{n}(t_n + k T_n), \\
 2 \pi f &= \dot{\phi}(t_n), \\
 T_n &= \ddot{\phi}^{-1/2}(t_n),
\end{align}
where 
\begin{align}
 \mathcal{H}_n (t) &= H_n(t) e^{-i n \phi_T(t)},
\end{align}
and the constants $b_k$ satisfy
\begin{align}
\sum_{k=-3}^3 b_k &= 1, \\
\sum_{k=1}^3 b_k \frac{k^{2p}}{(2p)!} &= \frac{(-i)^p}{2^{p+1} p!}, \quad 1 \leq p \leq 3, \\
b_{k} &= b_{-k}.
\end{align}

\section{\label{sec:results} Results}
\begin{figure*}
\includegraphics[width=\textwidth]{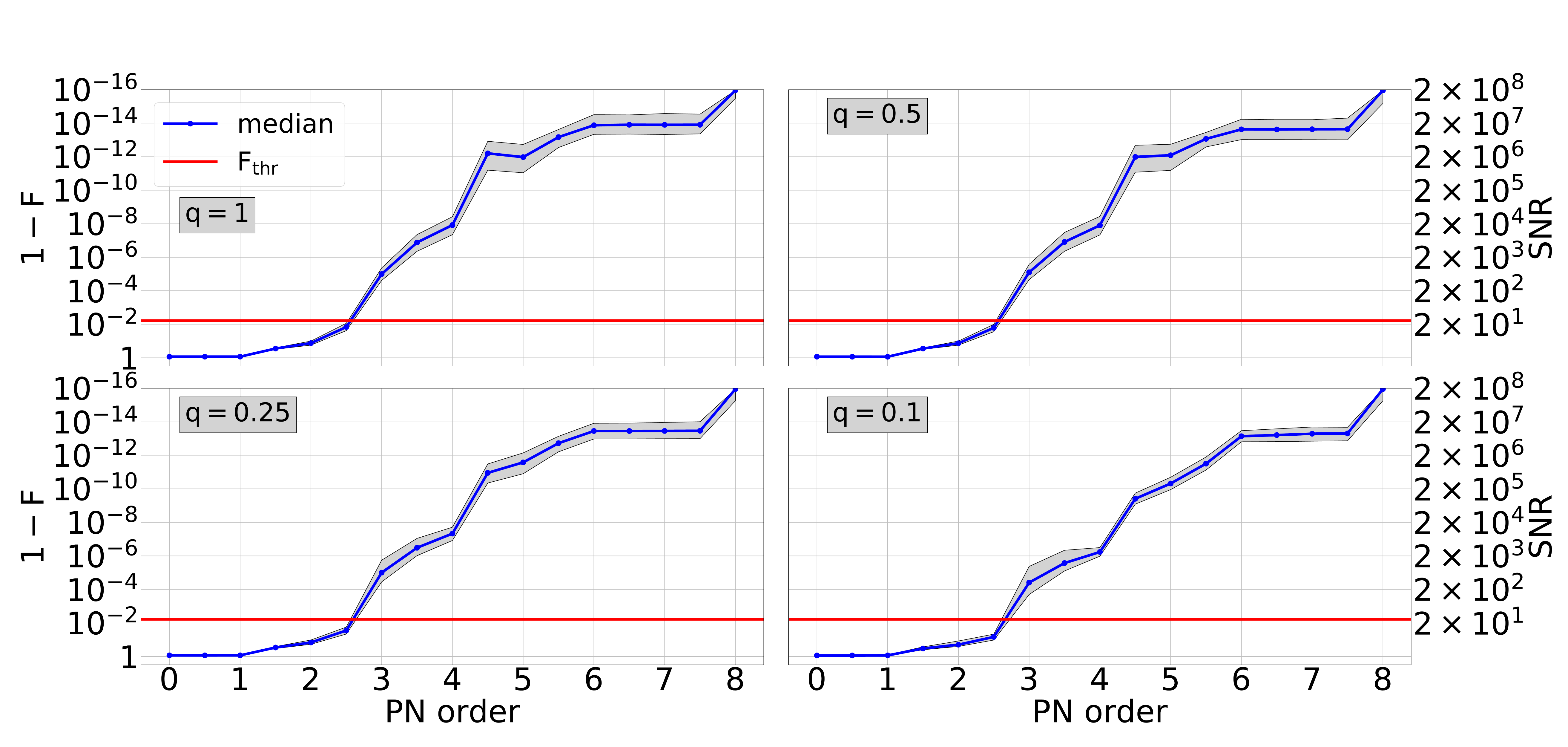}
\caption{\label{fig:faith_caseI} 
Results of the simulations for case \hyperref[sec:case(I)]{(I)}, with four different values of the binary mass ratio, i.e. $ q =1, 0.5, 0.25, 0.1$ as labeled in the panels. For each value of $q$, we show the unfaithfulness $(1-F)$ as a function of the waveform PN order. The blue line represents the median over $10^3$ simulated waveforms and the shaded area is limited by the 16\% and the 84\% quantiles (i.e. it corresponds to the $1\sigma$ confidence region). The red line is the threshold value $F = 0.994$, corresponding to a SBHB signal with $\text{SNR} = 25$ (on the right axes) and $D = 8$ waveform parameters.}
\end{figure*}

We run different simulations to verify the lowest PN order necessary to compute a faithful waveform. In all simulations we approximate the true signal $h_t$ with a waveform constructed with 8PN phasing (which we assume as our reference waveform), and we simulate signals $h$ at lower PN orders. Once both signals are obtained, we compute the faithfulness, $F$,  according to Eq.~\eqref{eq:faithfulness}, maximizing over the time and orbital phase. 
The resulting value is compared to a fiducial threshold of $F = 0.994$ corresponding to the situation where the intrinsic errors due to mismodeling are smaller than the statistical errors from the detector noise, for a waveform with a number of intrinsic parameters $D = 8$ and a signal with $\text{SNR}$ = 25 \cite{PhysRevD.95.104004}.  

We perform different simulations in order to fully characterize the faithfulness requirements for SBHB signals in LISA as a function of the relevant source parameters, namely:
\begin{enumerate}[(I)]
\item the mass ratio $q  =m_2 / m_1\leq1$. We extract $m_1$ from a log-flat distribution in $[10,100] \, \M_{\odot}$ and set different values of $q$, namely $q \in (0.1,\, 0.25,\, 0.5,\, 1)$ with $m_2 > 5 \, \M_{\odot}$;
\item the total mass $M  =m_1+m_2$. We fix $q=1$ and consider  binaries with different total masses, i.e. $M = 60, \, 120, \, 200, \,300 \, \M_{\odot}$;
\item the SBHB coalescence time, $t_c$. We again extract $m_1$ from a log-flat distribution  in the interval  $[10,100] \, \M_{\odot}$, with $q =1$ and generate SBHB waveforms starting 4 years to 150 years before merger (i.e. we simulate systems that at the start of the LISA mission are 4 years to 150 years before merger).
\end{enumerate}
In all cases, we consider spinning BHs, with spin magnitudes $ \chi_{1,2}$ extracted from a flat distribution $0 < \chi_{1,2} < 1$ and  directions uniformly distributed over a sphere. We further assume a 4-year  mission duration for LISA and quasi-circular orbits. Moreover, in cases (I) and (II), we consider SBHBs with coalescence time $t_c = 4 \yr$ (corresponding to a merger at the end of the LISA mission). We run  $N = 10^3$ random realizations to properly explore the parameter space. 

\begin{figure*}
\includegraphics[width=\textwidth]{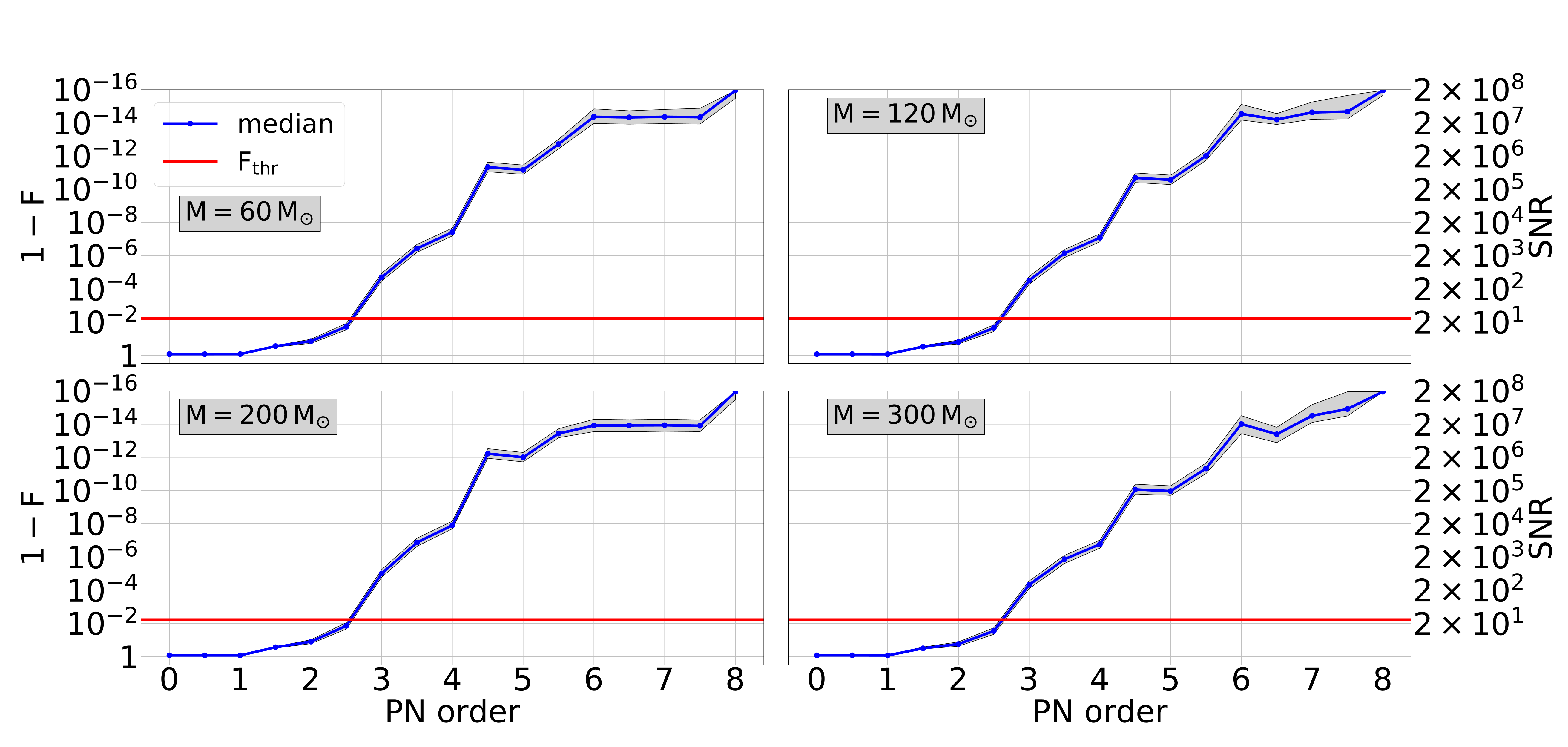}
\caption{\label{fig:faith_caseII} Same as Fig. \ref{fig:faith_caseI} for case \hyperref[sec:case(II)]{(II)}, i.e. for  different  values of the total binary mass, as reported in each panel and fixed $q = 1$. }
\end{figure*}

\subsection{\label{sec:case(I)} Case (I)}
In Fig. \ref{fig:faith_caseI} we present the results of our runs as a function of $q$.
The waveform is faithful at the adopted $F = 0.994$ threshold at PN orders larger than 3, almost irrespective of the value of $q$. In all cases, the inclusion of the $2.5$PN dissipative term only is insufficient to produce the faithful waveform for SBHBs merging in 4 years, and the inclusion of the 3PN term is crucial for an unbiased parameter estimation.
We notice that the convergence of the PN approximation is slower for lower mass ratios. This is a known characteristic of the PN approximation (see Fig. 1 in \cite{PhysRevD.85.064010}, or \cite{2003trso.conf..411B}), however it does not affect our conclusions.

\subsection{\label{sec:case(II)} Case (II)}
We now investigate the dependence of the faithfulness on the total mass of the SBHB.
In section \ref{sec:case(I)}, we have shown that our faithfulness requirement is only weakly sensitive on the mass ratio, so we fix $q=1$. 
More massive SBHBs reach coalescence at lower frequencies. Therefore, in the LISA band, these sources are detectable when they are closer to coalescence and we would expect to have to include higher PN orders to properly describe the signal, as relativistic effects  become  more important. However, for a fixed coalescence time more massive systems also emit at lower frequencies, therefore the number of cycles we need to match decreases. So even if the systems are slightly more relativistic, we have to match fewer cycles for the same observation time. 
Which one of these effects dominate the dependence of the PN requirements on the total mass will thus have to be determined from our simulations.

Our results are illustrated in Fig. \ref{fig:faith_caseII} for binaries with total mass from $M = 60 \, \M_{\odot}$ to $\M = 300 \, \M_{\odot}$. We observe that the requirement on the faithfulness is already satisfied by a 3PN waveform, even for the more massive binaries. Note that for a fixed PN order the faithfulness between the template  and  the reference waveform slightly decreases with the total binary mass. However, even if we were to observe systems close to coalescence for a long time, the faithfulness requirement would still be satisfied by a 3PN waveform, with a weak dependence from the total binary mass.

\subsection{\label{sec:case(III)} Case (III)}
Finally, we consider SBHBs with different coalescence times, $t_c$. In the previous cases, we fixed $t_{c} = 4 \yr$, however we expect these sources to only
represent  a small sub-sample of the population that LISA will observe.  Most of the SBHBs will be observed in their inspiral phase, several years before coalescence at a typical frequency of $10^{-2}\, \text{Hz}$ \cite{PhysRevLett.116.231102}. These sources will be slowly chirping in the LISA band and will not coalesce in the LIGO/Virgo band during the mission lifetime. Since these systems are far from coalescence, we expect lower PN order waveforms to be sufficient to properly describe and follow the signal.

\begin{figure*}
\includegraphics[width=\textwidth]{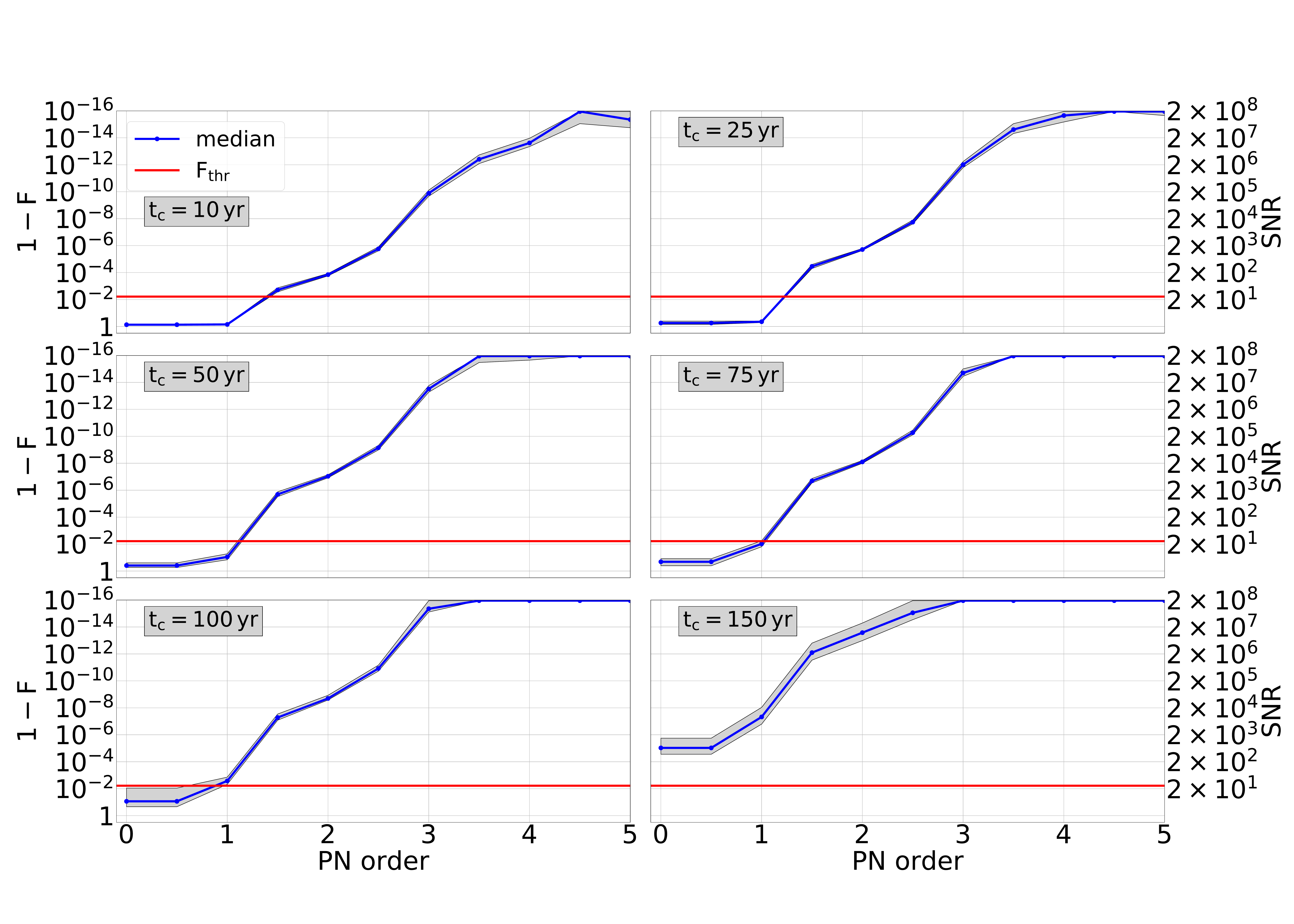}
\caption{\label{fig:faith_caseIII} Same as Fig. \ref{fig:faith_caseI} for case \hyperref[sec:case(I)]{(III)}, i.e. for different coalescence times, as reported in each panel.}
\end{figure*}

Since in the previous cases we found no strong dependence from the mass ratio and total binary mass, we choose to randomly draw $m_1$ from a log-flat distribution in $[10,100] \, \M_{\odot}$ and fix $q=1$. For each run, we fix different coalescence times, from $t_{c} = 10 \yr$ to $t_{c} = 150 \yr$. Results are shown in Fig. \ref{fig:faith_caseIII}.
SBHBs binaries with $t_{c} = 10, \, 25, \, 50, \, 75\yr$ can be properly described with just 1.5PN waveforms. Systems with $t_{c} = 100 \yr$ can be described accurately   with just a 1PN waveform and, finally, systems with $t_{c} = 150 \yr$ are well described by Newtonian (i.e. 0PN) waveforms. Since the PN accuracy requirement is 0PN for systems with $t_{c} = 150 \yr$, we conclude that also systems with longer merger time can be properly described by the same waveform accuracy.

In summary, we find that in order to guarantee that systematic waveform errors do not bias the recovery of SBHB parameters, 3PN templates are sufficient throughout the relevant parameter space.  From our analysis, the most important factor to determine the lowest PN corrections necessary to properly describe SBHB signals is the coalescence time. Even if the binary mass components and mass ratios  are expected to vary, their effect on waveform accuracy requirements is negligible.

\section{\label{sec:results_pop_model} Results from population models}
The goal of this study is to assess the waveform requirements for faithful signal reconstruction of the SBHBs detected by LISA. It is therefore useful to apply our analysis to expected SBHB populations, informed by the current LIGO/Virgo constraints.

We consider three population models, differing in the assumed SBHB mass function. In the first model, binary component masses are independently extracted from a log-flat distribution, with the only constrain that the total mass of the system has to lie in the range $[10\, \msun,100\, \msun]$, with $m_1, \, m_2 > 5 \, \M_{\odot}$ \cite{Abbott:2016nhf,Abbott:2016drs}. In the second and third model, the mass of the primary BH is obtained from a Salpeter distribution in the range $[5\,\msun,100\,\msun]$, and the secondary mass is drawn from a flat distribution in the range $[5\,\msun,m_1]$ and $[{\rm max}(5\,\msun,m_1/3),m_1]$ respectively. For each model we simulate 10 realizations of the expected SBHB population by drawing the local SBHB merger rate from the posterior distribution estimated by the LIGO O1 observations for either a log-flat or a Salpeter mass function, as reported in Figure 11 of \cite{PhysRevX.6.041015}. The merger rate is assumed to be constant in redshift (this does not have a major impact on LISA SBHB sources which are mainly at $z<0.5$ anyway). The population of SBHB observed by LISA as a function of mass, redshift and frequency is then computed as described in \cite{PhysRevLett.116.231102}, assuming circular binaries.

This procedure yields $N_{cat} = 30$ synthetic catalogs of SBHBs emitting in the LISA band. The GW signals from those systems are then integrated over the assumed 4 year LISA mission lifetime and the SNR is computed. We consider detectable, and thus retain, only  events with SNR$ >8$. The 30 catalogs yield an average number of $\approx 100$ SBHBs detectable by LISA above threshold. 

In this population analysis, we do not consider a single threshold value for the faithfulness as in Section \ref{sec:results}. Instead, we compute the faithfulness requirement for each event, based on its SNR, and we estimate the required PN order in the waveform model accordingly. In practice we proceed as follows:
\begin{itemize}
\item for each event we compute the SNR in the LISA detector (assuming a random sky location, inclination and polarization);
\item from the SNR we compute the faithfulness threshold from Eq.~\eqref{eq:expected_F};
\item we take the 8PN waveform as the true model signal and compute the faithfulness of waveforms at increasing PN orders until its value gets larger than the threshold.  
\end{itemize}
We repeat this procedure for all individual events with SNR$ >8$ in all 30 catalogues. We found similar results for the three population models. This had to be expected, since we have demonstrated in Section \ref{sec:results} that waveform requirements are largely independent of the mass and mass ratio of the SBHBs. Therefore, in the following we present results averaged over the 30 catalogues.

\begin{figure}
\includegraphics[width=0.5\textwidth]{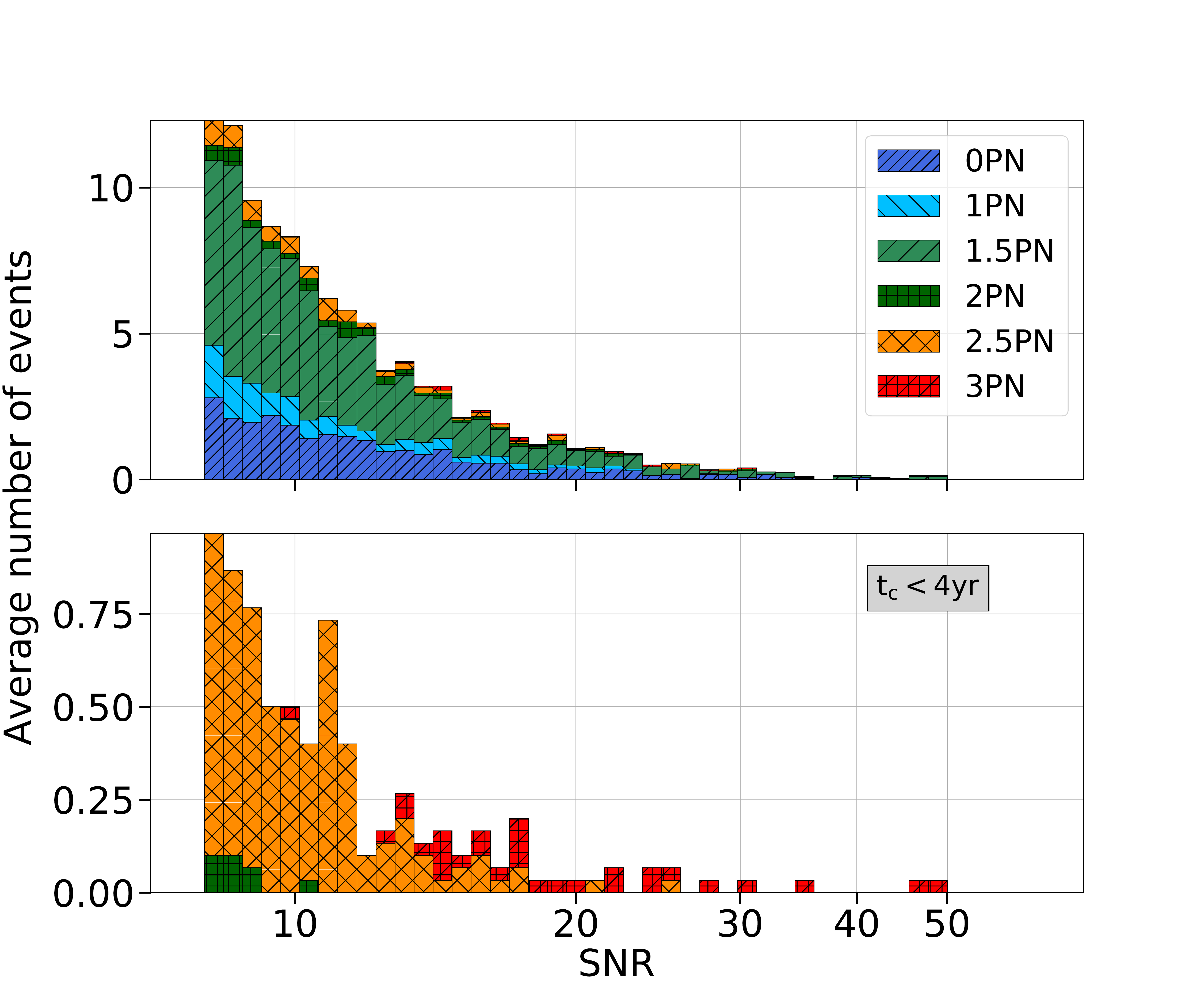}
\caption{\label{fig:faith_SNR_all_model} Top panel: stacked SNR distribution for PN sub-populations, labeled with different colors. Each bin is divided by the number of catalogs in our simulations, $N_{cat} = 30$, to provide mean estimates. Bottom panel:  same distribution but selecting only systems with $t_{c} < 4 \yr$. The PN sub-populations are constructed computing the faithfulness for each event and considering the lowest PN order satisfying the threshold value, from Eq.~\eqref{eq:expected_F}.}
\end{figure}

\begin{figure}
\includegraphics[width=0.5\textwidth]{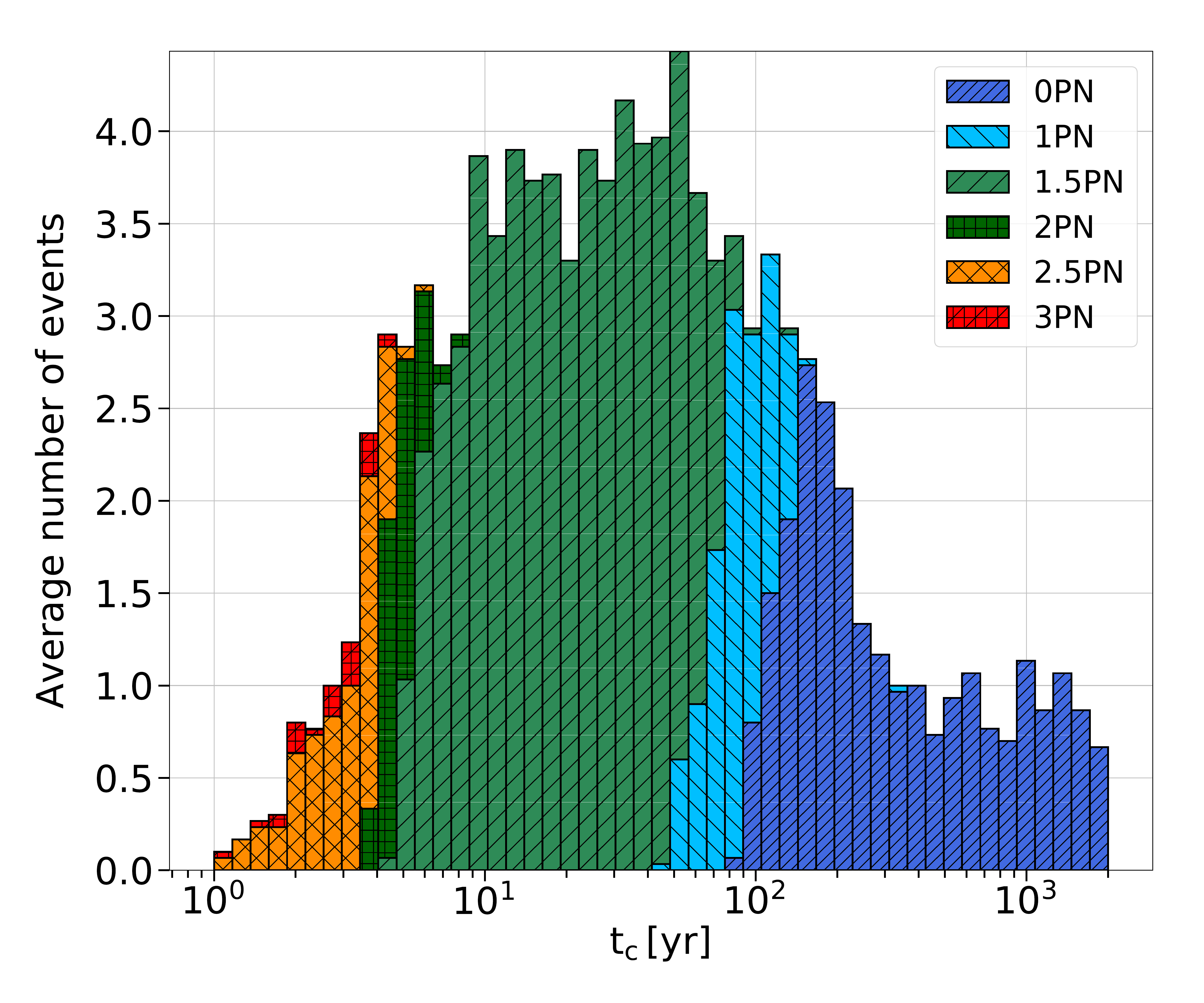}
\caption{\label{fig:tmerger_total_diff_pop_stacked} Stacked coalescence time distribution for each sub-population, color-coded as indicated in figure. }
\end{figure}

\begin{figure}\includegraphics[width=0.5\textwidth]{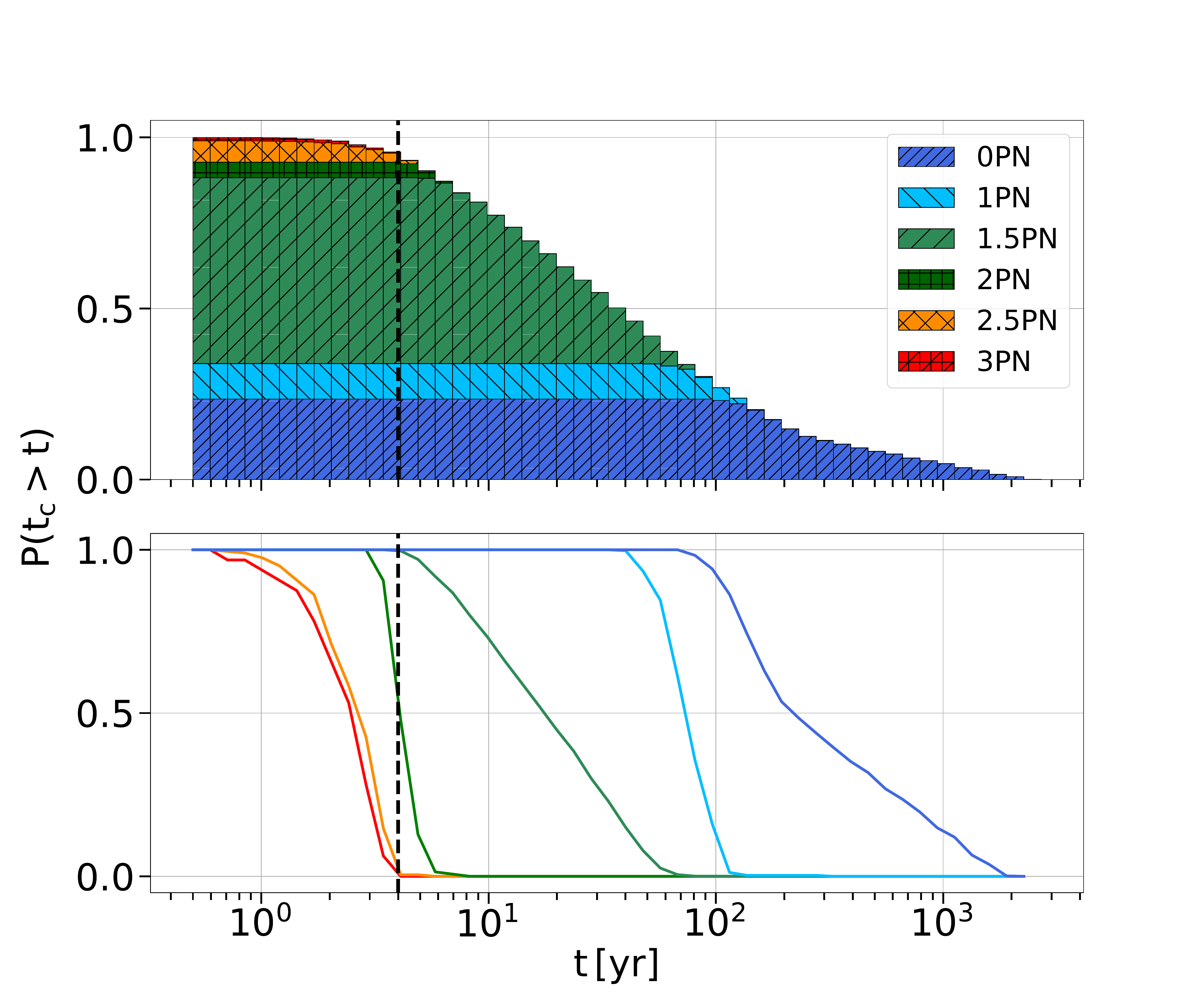}
\caption{\label{fig:tmerger_cumulative_two} Stacked cumulative coalescence time distribution for the complete catalogs (top panel) and for each PN sub-population (lower panel), color-coded as indicated in figure. The vertical dashed black line represents $t_c = 4 \yr$.}
\end{figure}

Fig. \ref{fig:faith_SNR_all_model} shows the average number of sources as a function of SNR, color-coded according to the PN accuracy requirement. The upper panel shows that for the majority of SBHBs, 1.5PN waveforms  are sufficient. Indeed, it is clear that  the vast majority ($\approx$90\%) of the SBHBs are described accurately  by using low PN waveform ($\PN < 2$) with only a small fraction of sources requiring 2.5PN or higher-PN  waveforms. The lower panel shows that the sub-population of SBHBs with $t_{c}  < 4 \yr$, i.e. those crossing to the LIGO/Virgo band within the LISA lifetime, generally require 2.5PN and 3PN waveforms. In fact 93\% (80\%) of the SBHBs requiring 3PN (2.5PN) waveforms fall in this sub-population. We also observe that the PN requirements are largely independent of the SNR for the global population, while for systems merging within the mission lifetime, more stringent PN requirements correspond to larger SNRs.

In Fig. \ref{fig:tmerger_total_diff_pop_stacked} we plot the average number of sources as a function of time to coalescence, again color-coded according to the PN order required for a faithful recovery of the signal. As expected from the analysis presented in the Section \ref{sec:results}, longer coalescence times imply less stringent requirement on the PN waveform accuracy necessary to track the signal phase. Indeed PN sub-populations are quite sharply separated in terms of coalescence time. For $t_{c} < 4 \yr$ the main contribution comes from the 2.5PN and 3PN sub-populations. In the interval $4 \yr < t_{c} < 10 \yr$, SBHBs require 2PN and 1.5PN corrections to be appropriately described. As can be seen in the top panel of Fig. \ref{fig:faith_SNR_all_model}, the larger sub-population is the 1.5PN one, which dominates the distribution for $t_{c} \in [10,100] \yr$, with an important contribution from the 1PN population for $t_{c} \simeq 100 \yr$. For longer $t_c$, most systems can be modeled by 0PN waveforms.

We also computed the cumulative distribution of sources as a function of coalescence time, divided according to the required PN order in waveform modeling. The result is presented in Fig. \ref{fig:tmerger_cumulative_two}. The upper panel shows the cumulative stacked distribution, color-coded for the different PN sub-populations. The lower panel shows the cumulative distribution separately for each PN sub-population. 
As expected from the previous results, all systems can be described using 3PN waveforms. Both the 3PN and 2.5PN sub-populations only contribute to the overall distribution for $t_{c} < 4 \yr$. The rest of the population (i.e. binaries with $t_{c} > 4 \yr$, accounting for $\sim 90\%$ of the overall sample) can be described by 2PN waveforms. About 75\% of the population lies at  $t_{c} > 10 \yr$ and requires 1.5PN waveform accuracy or lower; only 25\% of the detected SBHBs have $t_{c} > 100 \yr$ and can be mostly described by Newtonian waveforms.
In the lower panel, we can see the fraction of each sub-population at a given time. 
Again, the 3PN and 2.5PN sub-populations behave similarly and their support is almost completely inside $t_c < 4$~yrs. Similarly the 2PN sub-population shows support up to $t_{c} < 10 \yr$, while the 1.5PN one extends up to $t_{c} \simeq 70 \yr$. The 1PN and 0PN sub-populations show open support  up to $t_{c} \simeq 10^2 \yr$ and $t_{c} \simeq 10^3 \yr$ respectively.

These findings provide useful hints on the trade-off between the waveform accuracy necessary to follow the binary phase during the inspiral and the information recovered from the parameter estimation. Based on our simulations, we find that low PN templates can efficiently recover about 90\% of the signals allowing faithful and fast parameter estimation. However, for the multiband SBHBs merging within the mission lifetime, 3PN waveforms will be necessary to provide unbiased parameters and to precisely inform ground based detectors.

\section{\label{sec:discussion_conclusion}Discussion and Conclusion}
In this paper, we assumed circular orbits, since GW emission tends to circularize binary systems~\cite{PhysRev.136.B1224}. However, their residual eccentricity may not be negligible  \cite{PhysRevD.94.064020, Samsing:2018isx}, and therefore a study similar to ours has to be performed also for eccentric SBHBs. Moreover, from stellar evolution simulations we expect the presence of a mass gap in the black hole mass distribution \cite{Heger:2002by, Spera:2017fyx} between $[60,120] \, \M_{\odot}$, which we do not consider in this work. Nevertheless, this gap may be filled by ``second generation'' black holes in  cluster formation scenarios \cite{PhysRevD.95.124046,
PhysRevLett.120.151101}.

We determined the PN accuracy requirements necessary to properly track a SBHB signal during the inspiral phase in the LISA band. We show that for systems merging within the mission lifetime, a 3PN waveform is sufficient to perform unbiased parameter estimation, almost independently of  the mass ratio and the total binary mass. Longer coalescence times require less accurate templates to be properly modeled. In particular, assuming a 4 year LISA mission, systems with $t_c = 5, \,10, \, 50, \, 75 \yr$ require only a 1.5PN accurate waveform. Systems with $t_c = 100 \yr$ and $t_c = 150 \yr$ can be properly modeled by 1PN and Newtonian (i.e. 0PN) waveforms respectively. Spin effects and spin precession are found negligible as precession timescales are exceedingly long in these systems. 

We also tested our code on different SBHB population models. For each event predicted by these models, we have computed the corresponding PN accuracy requirement. Most of the detectable SBHBs will be properly described by low PN waveforms ($\PN \le 2$), while systems merging within the mission lifetime will require 2.5PN and 3PN waveform phasing. Indeed, the coalescence time distribution shows PN sub-populations reasonably separated in coalescence time, with systems requiring higher PN accuracy appearing only at small $t_c$. We  computed the cumulative distribution of the coalescence time for each PN sub-population. This shows in particular that for systems merging after the end of the LISA mission ($t_c > 4 \yr$) 2PN waveforms can be used to provide unbiased parameter estimation, with even lower PN orders needed as $t_c$ further increases.

Overall, our results provide an estimate of the waveform accuracy necessary to track the GW phase without compromising the recovered information. Even though all detectable SBHB signals can be properly described by 3PN waveforms, for a large fraction of sources 2PN waveforms alone will already satisfy the conditions for unbiased parameter estimation. Thus, our results may help the data analysis community in the construction of efficient and computationally viable algorithms.

\newpage
\acknowledgments
A.~M. and M.~C. acknowledge partial financial support from the INFN TEONGRAV specific initiative. A.~M. acknowledges networking support by the COST Action CA16104.
A.~K. and E.~B. acknowledge support from the H2020-MSCA-RISE-2015 Grant No. StronGrHEP-690904.
This work was supported by the Centre National d'{\'E}tudes Spatiales. A.~S. is supported by the Royal Society.

\bibliography{bibliography.bib}

\end{document}